\documentclass[aps,eqsecnum,preprint,floats,epsf,epsfig,nofootinbib]{revtex4}
\usepackage{epsfig}
\usepackage{graphicx}
\usepackage{subfigure}

\begin{document}
\def\be{\begin{eqnarray}}
\def\en{\end{eqnarray}}
\def\non{\nonumber}
\def\la{\langle}
\def\ra{\rangle}
\def\A{{\cal A}}
\def\B{{\cal B}}
\def\c{{\cal C}}
\def\d{{\cal D}}
\def\e{{\cal E}}
\def\p{{\cal P}}
\def\t{{\cal T}}
\def\nc{N_c^{\rm eff}}
\def\CP{{\it CP}~}
\def\CPP{{\it CP}}
\def\vp{\varepsilon}
\def\drho{\bar\rho}
\def\deta{\bar\eta}
\def\vma{{_{V-A}}}
\def\vpa{{_{V+A}}}
\def\J{{J/\psi}}
\def\ov{\overline}
\def\Bs{\overline B_s^0}
\def\Bd{\overline B_d^0}
\def\pr{{ Phys. Rev.}~}
\def\prl{{ Phys. Rev. Lett.}~}
\def\pl{{ Phys. Lett.}~}
\def\lsim{ {\ \lower-1.2pt\vbox{\hbox{\rlap{$<$}\lower5pt\vbox{\hbox{$\sim$}
}}}\ } }
\def\gsim{ {\ \lower-1.2pt\vbox{\hbox{\rlap{$>$}\lower5pt\vbox{\hbox{$\sim$}
}}}\ } }

\font\el=cmbx10 scaled \magstep2{\obeylines \hfill March, 2014}

\vskip 1.5 cm

\centerline{\large\bf Charmless Three-body Decays of $B_s$ Mesons}
\bigskip
\centerline{\bf Hai-Yang Cheng$^{1}$, Chun-Khiang Chua$^{2}$}
\medskip
\centerline{$^1$ Institute of Physics, Academia Sinica}
\centerline{Taipei, Taiwan 115, Republic of China}
\medskip
\centerline{$^2$ Department of Physics and Center for High Energy Physics}
\centerline{Chung Yuan Christian University}
\centerline{Chung-Li, Taiwan 320, Republic of China}

\medskip

\bigskip
\bigskip
\bigskip
\centerline{\bf Abstract}
\bigskip
\small

Branching fractions and {\it CP}-violating asymmetries of charmless $\Bs\to PPP$ decays with $P=K,\pi$ are studied using a simple model based on the factorization approach. The penguin-dominated modes $\Bs\to \ov K^0 K^+\pi^-$ and $K^0K^-\pi^+$ have the largest rates among the three-body $B_s$ decays and they are dominated by the $K_0^*(1430)$ resonances and nonresonant contributions.
The branching fraction of $\Bs\to K^0K^+K^-$ is predicted to be of order
$1.4\times 10^{-6}$, which is in the middle of the measured result  $(0.2-3.4)\times 10^{-6}$ obtained by LHCb. We find sizable \CP violation in $K^0\pi^+\pi^-$, $K^0\pi^0\pi^0$, $K^0K^+K^-$ and $K_SK_SK_S$ channels. Just as the $B_u$ sector, the \CP asymmetries  of  $\Bs\to K^0\pi^+\pi^-$ and $\Bs\to K^0K^+K^-$ have similar magnitudes but are opposite in sign.
Several model-independent relations based on $U$-spin symmetry for \CP asymmetries are derived. Although they cannot be tested by the present available data, they can be checked by the dynamical calculations. Since the $U$-spin symmetry which relates various matrix elements of scalar densities, for example, $\la K^0K|\bar dq|0\ra$ and  $\la \ov K^0\pi|\bar sq|0\ra$, is badly broken, the derived $U$-spin relations are generally not well respected.

\pagebreak

\section{Introduction}
Recently LHCb has made the first observation of charmless three-body decays of the $B_s^0$ meson with a $K_S^0$ meson in the final state \cite{Aaij:Bs}. Specifically, LHCb has obtained the following branching fractions
\be \label{eq:Bsexpt}
 \B(B_s^0\to \buildrel (-)\over {K^0} K^\pm\pi^\mp) &=& (73.6\pm5.7\pm6.9\pm3.0)\times 10^{-6}, \non \\
 \B(B_s^0\to \ov K^0\pi^+\pi^-) &=& (14.3\pm2.8\pm1.8\pm0.6)\times 10^{-6}, \non \\
  \B(B_s^0\to \ov K^0 K^+K^-) &\in & [0.2;3.4]\times 10^{-6}.
\en
The first observation of the decay modes $B_s\to K_S\pi^+\pi^-$ and $B_s\to K_SK^\pm\pi^\mp$ is an important step at the LHCb towards extracting information on the mixing-induced {\it CP}-violating phase in the $B_s^0$ system and the weak phase $\gamma$ from these decays.

In the charged $B$ meson sector, LHCb recently found evidences of direct \CP violation in charmless three-body decays: $B^+\to\pi^+\pi^+\pi^-$, $B^+\to K^+K^+K^-$ and $B^+\to K^+K^-\pi^+$ \cite{LHCb:Kppippim,LHCb:pippippim}. Moreover, LHCb has also observed large asymmetries in some localized regions of phase space. For example, the regional \CP asymmetry in $B^+\to\pi^+\pi^+\pi^-$ is of order 58\% for $m^2_{\pi^-\pi^- \rm ~low}<0.4$ GeV$^2$ and $m^2_{\pi^+\pi^- \rm ~high}>15$ GeV$^2$. Hence, significant signatures of \CP violation were found in the  low mass regions devoid of most of the known resonances. LHCb also discovered interesting correlations of the \CP violation between the decays, namely,
\be \label{eq:CPcorrelation}
\A_{CP}(B^-\to\pi^-K^+K^-)&\approx& -\A_{CP}(B^-\to\pi^-\pi^+\pi^-) \non \\
\A_{CP}(B^-\to K^-\pi^+\pi^-) &\approx& -\A_{CP}(B^-\to K^-K^+K^-).
\en
It will be interesting to see if the same pattern of \CP violation occurs in $B_s^0$ meson decays.

Contrary to $B_u$ and $B_d$ mesons,  for three-body $B_s$ decays, the $b\to sq\bar q$ penguin transitions
contribute to the final states with even number of kaons, namely,
$KK\pi$, while $b\to uq\bar q$ tree and $b\to dq\bar
q$ penguin transitions contribute to final states with odd number
of kaons, e.g. $K\pi\pi$ and $KKK$. If the final-state pseudoscalar meson is restricted to be a kaon or a pion, then the allowed penguin-dominated three-body decays are $\ov B_s^0\to K^0 K^-\pi^+, \ov K^0 K^+\pi^-, K^+K^-\pi^0, K^0\ov K^0\pi^0$. Therefore, it is expected that the first two modes have the largest rates among three-body $B_s$ decays.
Tree-dominated decays are $\Bs\to K^+\pi^-\pi^0,K^0\pi^+\pi^-,K^0\pi^0\pi^0$ and $\Bs\to K^+K^-K^0,\ov K^0K^0K^0$.

This work is organized as follows. We outline the framework in Sec. II for the evaluation of the resonant and nonresonant contributions to three-body decays. Before proceeding to the numerical results, we discuss in Sec. III some model-independent flavor symmetry relations in which some of $B_s\to PPP$ decays can be related to $B_d\to PPP$ ones by $U$-spin symmetry. The calculated branching fractions and \CP violation are presented in Sec. IV. Conclusions are given in Sec. V. Factorizable amplitudes of various $\Bs\to PPP$ decays are shown in the appendix.

\section{Framework}
The evaluation of three-body decays of $B_{u,d}$  mesons has been studied in details in \cite{CC:3body}. We shall follow the same framework based on the factorization approach to discuss the charmless 3-body decays of the $B_s$ meson. Here we shall recapitulate the main points.

Three-body decays of heavy mesons receive both resonant and nonresonant
contributions. Consider the charmless three-body $B_s$ decay $\Bs\to P_1P_2P_3$.
 Under the factorization hypothesis, its decay amplitude
consists of three distinct factorizable terms: (i) the current-induced process with a meson
emission, $\la\Bs\to P_1P_2\ra\times \la 0\to P_3\ra$,
(ii) the transition process,  $\la\Bs\to P_3\ra\times \la
0\to P_1P_2\ra$, and (iii) the annihilation process $\la
\Bs\to 0\ra\times \la 0\to P_1P_2P_3\ra$, where $\la A\to
B\ra$ denotes a $A\to B$ transition matrix element.

\subsection{Nonresonant contributions}
Dominant nonresonant contributions to tree-dominated three-body decays arise
from the $b\to u$ tree transition $\la\Bs\to P_1P_2\ra\times \la 0\to P_3\ra$ which can be evaluated using heavy meson chiral perturbation theory (HMChPT) \cite{LLW}. Take the decay $\Bs\to K^0 K^-\pi^+$ as an example for illustration. Its factorizable amplitude is given in Eq. (\ref{eq:AK0Kppim}). We first focus on nonresonant contributions.
The $b\to u$ transition amplitude evaluated using HMChPT reads \cite{LLW}
\be
 A_{\rm current-ind}^{\rm HMChPT} &\equiv&\la \pi^+(p_2) K^0(p_3)|(\bar u b)_{V-A}|\ov B^0_s\ra^{NR} \la K^-(p_1)|(\bar s u)_{V-A}|0\ra \non \\
&=&  -\frac{f_K}{2}\left[2 m_1^2 r+(m_{B_s}^2-s_{23}-m_1^2) \omega_+
 +(s_{12}-s_{13}-m_2^2+m_3^2) \omega_-\right],
\en
with the form factors
 \be \label{eq:r&omega}
 \omega_+ &=& -{g\over f_\pi^2}\,{f_{B^*}m_{B^*}\sqrt{m_{B_s}m_{B^*}}\over
 s_{12}-m_{B^*}^2}\left[1-{(p_{_{B_s}}-p_3)\cdot p_3\over
 m_{B^*}^2}\right]+{f_{B_s}\over 2f_\pi^2}, \non \\
 \omega_- &=& {g\over f_\pi^2}\,{f_{B^*}m_{B^*}\sqrt{m_{B_s}m_{B^*}}\over
 s_{12}-m_{B^*}^2}\left[1+{(p_{_{B_s}}-p_3)\cdot p_3\over
 m_{B^*}^2}\right], \non \\
 r &=& {f_{B_s}\over 2f_\pi^2}-{f_{B_s}\over
 f_\pi^2}\,{p_{B_s}\cdot(p_2-p_3)\over
 (p_{_{B_s}}-p_2-p_3)^2-m_{B_s}^2}+{2gf_{B^*}\over f_\pi^2}\sqrt{m_{B_s}\over
 m_{B^*}}\,{(p_{_{B_s}}-p_3)\cdot p_3\over s_{12}-m_{B^*}^2} \non \\
 &-& {4g^2f_{B_s}\over f_\pi^2}\,{m_{B_s}m_{B^*}\over
 (p_{_{B_s}}-p_2-p_3)^2-m_{B_s}^2}\,{p_3\!\cdot\!p_2-p_3\!\cdot\!(p_{_{B_s}}-p_3)\,p_2\!\cdot\!
 (p_{_{B_s}}-p_3)/m_{B^*}^2 \over s_{12}-m_{B^*}^2 },
 \en
where $s_{ij}\equiv (p_i+p_j)^2$. Likewise,
\be
 &&\la K^+(p_1)\pi^-(p_2)|(\bar d b)_{V-A}|\ov B^0_s\ra^{NR} \la \ov K^0(p_3)|(\bar s d)_{V-A}|0\ra \non \\
&=&  -\frac{f_K}{2}\left[2 m_3^2 r'+(m_{B_s}^2-s_{12}-m_3^2) \omega'_+
 +(s_{23}-s_{13}-m_2^2+m_1^2) \omega'_-\right]
\en
for the decay $\Bs\to \ov K^0 K^+\pi^-$,
where the form factors $\omega'_\pm$ and $r'$ are obtained from $\omega_\pm$ and $r$, respectively, with the replacement $p_1\leftrightarrow p_3$.

However, the predicted nonresonant rates due to $B\to P_1P_2$ transition alone already exceed the measured total branching fractions for the tree-dominated modes e.g. $\Bd\to\pi^-\pi^+\pi^-$, $\Bd\to\pi^-K^+K^-$ and $\Bs\to K^0\pi^+\pi^-$. For example, the nonresonant branching fraction
of the last decay channel estimated using HMChPT is found to be of order $32\times 10^{-6}$, which is even larger than the total branching fraction of order $14\times 10^{-6}$ [see Eq. (\ref{eq:Bsexpt})].  The issue has to do with the applicability of HMChPT. When it is applied to three-body decays, two of the final-state pseudoscalars  have to be soft. If the soft meson result is assumed to be the same in the whole Dalitz plot, the decay rate will be
greatly overestimated. To overcome this issue, we have proposed in \cite{CCS:nonres} to parameterize the momentum dependence of nonresonant amplitudes induced by $b\to u$ transition in an exponential form
\be \label{eq:ADalitz}
  A_{\rm current-ind}=A_{\rm current-ind}^{\rm
  HMChPT}\,e^{-\alpha_{_{\rm NR}}
p_B\cdot(p_1+p_2)}e^{i\phi_{12}},
\en
so that the HMChPT results are recovered in the soft pseudoscalar meson limit.
The parameter $\alpha_{_{\rm NR}}=0.081^{+0.015}_{-0.009}\,{\rm GeV}^{-2}$ \cite{CC:3body} is fixed by the measured
nonresonant rate in the tree-dominated decay $B^-\to\pi^+\pi^-\pi^-$,
where the nonresonant background due to the penguin diagram is suppressed by the smallness of penguin Wilson coefficients.

The other two types of nonresonant contributions to $\Bs\to K^0 K^-\pi^+$ are
 \be \label{eq:KpiBpi}
&& \la \pi^+(p_1)K^-(p_2)|(\bar s d)_{V-A}|0\ra^{NR}
\la K^0(p_3)|(\bar d b)_{V-A}|\Bs \ra  \non\\
&=& F_1^{K\pi}(s_{12})F_1^{B_sK}(s_{12})\left[s_{23}-s_{13}-{(m_{B_s}^2-m_\pi^2)(m_K^2-m_\pi^2)
\over s_{12}}\right] \non
\\ &+&  F_0^{K\pi}(s_{12})F_0^{B_sK}(s_{12}){(m_{B_s}^2-m_\pi^2)(m_K^2-m_\pi^2)
\over s_{12}}
\en
and
\be
\la \pi^+(p_1)K^-(p_2)|\bar sd|0\ra ^{NR}\la K^0(p_3)|\bar db|\Bs \ra = {m_{B_s}^2-m_K^2\over m_b-m_d}\,F_0^{B_sK}(s_{12})\la\pi^+(p_1)K^-(p_2)|\bar sd|0\ra ^{NR}.
\en
As stressed in \cite{CC:3body,CCS:nonres}, the matrix element of scalar densities $\la M_1M_2|\bar q_1q_2|0\ra$ must have a large nonresonant component in order to explain the large nonresonant signals observed in penguin-dominated three-body $B$ decays. In other words, nonresonant contributions to penguin-dominated three-body $B$ decays are also penguin dominated. From the study of $B_{u,d}\to KKK$ decays, it was found \cite{CC:3body}
\be \label{eq:KKssme}
 \la K^-(p_1) K^+(p_2)|\bar s s|0\ra
 =\frac{v}{3}(3 F_{NR}+2F'_{NR})+\sigma_{_{\rm NR}}
 e^{-\alpha s_{12}},
\en
where the parameters $v$, $F_{\rm NR}$, $F'_{\rm NR}$ and $\sigma_{_{\rm NR}}$ are defined in \cite{CC:3body} and $\alpha=(0.14\pm0.02){\rm GeV}^{-2}$. However, if SU(3) symmetry is applied to the matrix element of scalar densities so that $\la K^-(p_1)\pi^+(p_2)|\bar sd|0\ra^{NR}=\la K^-(p_1)K^+(p_2)|\bar ss|0\ra^{NR}$, we found in \cite{CC:3body} that the predicted $\A_{CP}(B^-\to\pi^-K^+K^-)$ and $\A_{CP}(B^-\to K^-\pi^+\pi^-)$ are wrong in signs when confronted with experiment. The correlations of the \CP violation between the charged $B$ decays shown in Eq. (\ref{eq:CPcorrelation}) have led to the conjecture that $\pi^+\pi^-\leftrightarrow K^+K^-$
rescattering may play an important role in the generation of the strong phase
difference needed for such a violation to occur.
It is thus plausible that a strong phase in  $\la K^-\pi^+|\bar sd|0\ra$ induced from final-state interactions might flip the sign of \CP asymmetry. A fit to the data of $\Bd\to K^-\pi^+\pi^-$ yields \cite{CC:3body}
\be \label{eq:KpimeNew1}
 \la K^-(p_1)\pi^+(p_2)|\bar s d|0\ra^{\rm NR}
 &\approx& \frac{v}{3}(3 F_{\rm NR}+2F'_{\rm NR})+\sigma_{_{\rm NR}}
 e^{-\alpha s_{12}}e^{i\pi}\left(1+4{m_K^2-m_\pi^2\over s_{12}}\right),
\en
with a strong phase of order $180^\circ$.

\subsection{Resonant contributions}
Resonant effects are commonly described in terms of the usual Breit-Wigner
formalism. Contributions of vector meson and scalar resonances to the three-body and two-body matrix elements are given by \cite{CC:3body}
\be
 \la \pi^+(p_1)K^-(p_2)|(\bar sb)_\vma|\ov B^0\ra^R &=& \sum_i
{g^{K_i^*\to K^-\pi^+}\over s_{12}-
m_{K_i^*}^2+im_{K_i^*}\Gamma_{K_i^*}}\sum_{\rm
pol}\vp^*\cdot
(p_1-p_2)\la \ov K^{*0}_i|(\bar sb)_\vma|\ov B^0\ra \non \\
&&-
{g^{K_{0}^*\to K^-\pi^+}\over s_{12}-
m_{K_{0}^*}^2+im_{K_{0}^*}\Gamma_{K_{0}^*}}\la \ov K^{*0}_{0}|(\bar sb)_\vma|\ov B^0\ra, \non \\
\la \pi^+(p_1)K^-(p_2)|(\bar sd)_\vma|0\ra^R &=& \sum_i
{g^{K^*_i\to K^-\pi^+}\over
s_{12}-m_{K^*_i}^2+im_{K^*_i}\Gamma_{K^*_i}}\sum_{\rm
pol}\vp^*\cdot
(p_1-p_2)\la \ov K^{*0}_i|(\bar sd)_\vma|0\ra \non \\
&&- {f_{{K^*_0}} g^{{K^*_{0}}\to K^-\pi^+}\over
s_{12}- m_{K^*_{0}}^2+im_{K^*_{0}}\Gamma_{K^*_{0}}}(p_1+p_2)_\mu,  \\
 \la \pi^+(p_1)K^-(p_2)|\bar s d|0\ra^R
 &=& -\frac{m_{{K^*_0}} \bar f_{{K^*_0}} g^{{K^*_0}\to K^-\pi^+}}{s_{12}- m_{{K^*_0}}^2+i
 m_{{K^*_0}}\Gamma_{{K^*_0}}}, \non
 \en
with $K_i^*=K^*(892), K^*(1410),K^*(1680),\cdots$, and $K_{0}^*=K_0^*(1430)$.
In the above equations we have two different types of decay constants for the scale meson $K_0^*$ defined by $\la K_0^*|\bar sd|0\ra=\bar f_{K^*_0} m_{K^*_0}$ and  $\la K_0^*(p)|\bar s\gamma_\mu d|0\ra=f_{K^*_0}p_\mu$. They are related via
\be
\bar f_{K^*_0} = {m_{K^*_0}\over m_s(\mu)-m_d(\mu)}f_{K^*_0}.
\en
Their values are given in \cite{CCY:SP}.

\section{$U$-spin symmetry}
Before proceeding to the numerical results in the next section, here
we would like to discuss some model-independent flavor symmetry relations in which
some of $B_s\to PPP$ decays can be related to $B_d\to PPP$ ones via $U$-spin symmetry. Hence these  relations can be used to cross-check the dynamical calculations. As an example, the decay amplitudes of $\Bs\to K^0 \pi^+\pi^-$ and $\Bd\to \ov K^0 K^+K^-$ can be related to each other in the limit of $U$-spin symmetry:
\be
A(\Bs\to K^0 \pi^+\pi^-) &=& V_{ub}^*V_{ud}\la K^0 \pi^+\pi^-|O_d^u| \Bs \ra+V_{cb}^*V_{cd}\la K^0\pi^+\pi^-|O_d^c|\Bs\ra, \non \\
A( \Bd\to \ov K^0 K^+K^-) &=& V_{ub}^*V_{us}\la \ov K^0 K^+K^-|O_s^u| \Bd \ra+V_{cb}^*V_{cs}\la \ov K^0 K^+K^-|O_s^c|\Bd\ra,
\en
where the 4-quark operator $O_s$ is for the $b\to sq_1\bar q_2 $ transition and $O_d$ for the $b\to dq_1\bar q_2$ transition. The assumption of
$U$-spin symmetry implies that under $d \leftrightarrow s$ transitions
\be
\la \ov K^0K^+K^-|O_s^u|\Bd\ra=\la K^0 \pi^+\pi^-|O_d^u|\Bs\ra, \qquad
\la \ov K^0K^+K^-|O_s^c|\Bd\ra=\la K^0 \pi^+\pi^-|O_d^c|\Bs\ra,
\en
which can be checked from Eq. (\ref{eq:AK0pippim}) for $\Bs\to K^0\pi^+\pi^-$ and  Eq. (A4) of \cite{CCS:nonres} for $\Bd\to\ov K^0K^+K^-$.
Using the relation for the CKM matrix \cite{Jarlskog}
\be
{\rm Im}(V_{ub}^*V_{ud}V_{cb}V_{cd}^*)=-{\rm Im}(V_{ub}^*V_{us}V_{cb}V_{cs}^*),
\en
it is straightforward to show that
\be
&& |A(\Bd\to \ov K^0K^+K^-)|^2-|A(B_d^0\to K^0 K^-K^+)|^2 \non \\
&& =|A(\Bs\to K^0 \pi^+\pi^-)|^2-|A(B_s^0\to \ov K^0\pi^-\pi^+)|^2.
\en
Hence, $U$-spin symmetry leads to the relation
\be
{\A_{CP}(\Bs\to K^0 \pi^+\pi^-)\over  \A_{CP}(\Bd\to \ov K^0 K^+K^-)} &=& -{\Gamma(\Bd\to \ov K^0K^+K^-)\over \Gamma(\Bs\to K^0\pi^+\pi^-)}.
\en
Therefore, we have the following $U$-spin symmetry relations \cite{London}
\be \label{eq:Uspin}
\A_{CP}(\ov B_s^0\to K^0 K^- \pi^+) &=& -\A_{CP}(\ov B^0_d\to \ov K^0 K^+ \pi^-)\,{\B(\ov B^0_d\to \ov K^0 K^+ \pi^-)\over \B(\ov B_s^0\to K^0 K^- \pi^+)} \,{\tau(B_s^0)\over \tau(B_d^0)}, \non  \\
\A_{CP}(\ov B_s^0\to \ov K^0 K^+ \pi^-) &=& -\A_{CP}(\ov B^0_d\to K^0 K^- \pi^+)\,{\B(\ov B^0_d\to K^0 K^- \pi^+)\over \B(\ov B_s^0\to \ov K^0 K^+ \pi^-)} \,{\tau(B_s^0)\over \tau(B_d^0)}, \non  \\
\A_{CP}(\ov B_s^0\to K^0\pi^+\pi^-) &=& -\A_{CP}(\ov B^0_d\to \ov K^0K^+K^-)\,{\B(\ov B^0_d\to \ov K^0K^+K^-)\over \B(\ov B_s^0\to K^0\pi^+\pi^-)}\, {\tau(B_s^0)\over \tau(B_d^0)},  \\
\A_{CP}(\ov B_s^0\to K^0K^+K^-) &=& -\A_{CP}(\ov B^0_d\to  \ov K^0\pi^+\pi^-)\,{\B(\ov B^0_d\to \ov K^0\pi^+\pi^-)\over \B(\ov B_s^0\to K^0K^+K^-)}\, {\tau(B_s^0)\over \tau(B_d^0)}, \non \\
\A_{CP}(\ov B_s^0\to K^0K^0\ov K^0) &=& -\A_{CP}(\ov B^0_d\to  \ov K^0\ov K^0 K^0)\,{\B(\ov B^0_d\to \ov K^0\ov K^0K^0)\over \B(\ov B_s^0\to K^0K^0\ov K^0)}\, {\tau(B_s^0)\over \tau(B_d^0)}. \non
\en
Unlike the $U$-spin relation in two-body decays, namely \cite{He},
\be \label{eq:Uspin1}
\A_{CP}(\Bs\to K^+\pi^-)=-\A_{CP}(\Bd\to K^-\pi^+)\,{\B(\Bd\to K^-\pi^+)\over \B(\Bs\to K^+\pi^-)}\,{\tau(B_s^0)\over \tau(B_d^0)},
\en
which has been well tested with the recent LHCb measurement of \CP violation in $\Bs\to K^+\pi^-$ \cite{LHCb:Bs}, the relations in Eq. (\ref{eq:Uspin}) cannot be checked by the present available data. Nevertheless, they can be tested by our dynamical calculations as shown below.

\section{Branching fractions and \CP asymmetries}

For numerical calculations, we shall use the input parameters  given in \cite{CC:3body} except the form factor $F_0^{B_sK}$.
As discussed in \cite{CC:Bs},
this form factor at $q^2=0$ calculated in the literature ranges from 0.23 to 0.31\,. We find that the form factor $F_0^{B_sK}(0)\approx 0.31$ gives a better description of the measured branching fractions of $\ov B_s^0\to \buildrel (-)\over {K^0} K^\mp\pi^\pm$ and $K^0\pi^+\pi^-$.

The calculated branching fractions and \CP asymmetries are summarized in Table \ref{tab:BsBRCP}. The theoretical errors shown there are
from the uncertainties in (i) the parameter $\alpha_{_{\rm NR}}$ appearing in Eq. (\ref{eq:KKssme})
which governs the momentum dependence of the nonresonant
amplitude, (ii) the strange quark mass $m_s$ for decay modes involving kaon(s), the form factor
$F^{B_sK}_0$  with the uncertainty assigned to be $0.03$ and the nonresonant parameter $\sigma_{_{\rm NR}}$ given after  Eq. (\ref{eq:KKssme}), and
(iii) the unitarity angle $\gamma$.

\begin{table}[tb]
 \caption{Branching fractions (in units of $10^{-6}$) and direct \CP asymmetries (in \%) in $\Bs\to PPP$ decays. Experimental results of branching fractions are taken from \cite{Aaij:Bs}. Theoretical errors correspond to the uncertainties in (i)
$\alpha_{_{\rm NR}}$, (ii) $F^{B_sK}_0$, $\sigma_{_{\rm
NR}}$ and $m_s(\mu)=(90\pm 20) $MeV at $\mu=2.1$ GeV,  and (iii) $\gamma=(69.7^{+1.3}_{-2.8})^\circ$.
 } \label{tab:BsBRCP}
\begin{ruledtabular}
 \begin{tabular}{l c c  r }
 {Modes} & $\B_{\rm theory}$
   &  $\B_{\rm expt}$ & $\A_{CP}(\%)$   \\  \hline
   $\ov K^0 K^+\pi^-$ & $35.3^{+0.3+15.7+0.0}_{-0.2-~9.8-0.0}$ & $$ & $-1.9^{+0.1+0.1+0.0}_{-0.1-0.1-0.0}$ \\
   $K^0 K^-\pi^+$ & $36.7^{+0.2+14.9+0.1}_{-0.2-~9.0-0.1}$ & $$ & $4.6^{+0.3+1.1+0.1}_{-0.3-1.1-0.1}$ \\
   $\buildrel (-)\over {K^0} K^\mp\pi^\pm$ & $72.0^{+0.4+21.6+0.1}_{-0.2-13.3-0.1}$ & $73.6\pm5.7\pm6.9\pm3.0$ & \\
   $K^+K^-\pi^0$ & $19.1^{+0.0+7.5+0.0}_{-0.0-4.5-0.0}$ & & $3.3^{+0.0+1.4+0.0}_{-0.0-1.5-0.1}$ \\
   $K^0\ov K^0\pi^0$ & $20.3^{+0.3+8.7+0.0}_{-0.4-5.5-0.0}$ & & $0.8^{+0.0+0.1+0.0}_{-0.0-0.1-0.0}$ \\
   $K^0\pi^+\pi^-$ & $12.7^{+0.5+0.5+0.1}_{-0.5-0.3-0.1}$ & $14.3\pm2.8\pm1.8\pm0.6$ & $8.0^{+0.9+1.1+0.0}_{-1.4-1.3-0.1}$ \\
   $K^+\pi^-\pi^0$ & $16.9^{+0.4+2.1+0.0}_{-0.4-1.9-0.0}$ & &
   $0.6^{+0.3+0.3+0.0}_{-0.5-0.3-0.0}$ \\
   $K^0\pi^0\pi^0$ & $0.48^{+0.01+0.06+0.01}_{-0.01-0.06-0.01}$ & & $-33.1^{+5.6+0.6+0.1}_{-3.9-0.6-0.1}$ \\
   $K^0K^+K^-$ & $1.4^{+0.0+0.3+0.2}_{-0.0-0.1-0.2}$ & $\in$ [0.2;~3.4] & $-17.4^{+0.1+0.7+0.4}_{-0.2-0.5-0.4}$ \\
   $K_SK_SK_S$ & $0.22^{+0.00+0.07+0.01}_{-0.00-0.10-0.01}$ & & $-13.4^{+0.1+0.4+0.2}_{-0.1-0.4-0.2}$
 \end{tabular}
 \end{ruledtabular}
 \end{table}

\begin{table}[t]
\caption{Predicted branching fractions (in units of $10^{-6}$) of resonant and
nonresonant (NR) contributions to $\Bs\to
\ov K^0 K^+\pi^-$ and $K^0K^-\pi^+$. }
\begin{ruledtabular} \label{tab:BsKKpi}
\begin{tabular}{l r |l r }
 $\Bs\to\ov K^0 K^+\pi^-$ & & $\Bs\to K^0 K^-\pi^+$ & \\ \hline
% Decay mode &  &Decay mode &  \\ \hline
 $K^{*0}\ov K^0$ & \qquad\quad $1.5^{+0.0+2.4+0.0}_{-0.0-0.9-0.0}$ ~~~  &
 $\ov K^{*0}K^0$ & $3.8^{+0.0+0.8+0.0}_{-0.0-0.7-0.0}$ \\
 $K^{*-}K^+$ & \qquad\quad $3.5^{+0.0+0.7+0.1}_{-0.0-0.7-0.1}$ ~~~ & $K^{*+}K^-$ & $2.6^{+0.0+2.7+0.1}_{-0.0-1.1-0.1}$ \\
 $K_0^{*0}(1430)\ov K^0$ & \qquad\quad $0.6^{+0.0+0.9+0.0}_{-0.0-0.4-0.0}$ ~~~ & $\ov K_0^{*0}(1430)K^0$ & $14.5^{+0.0+3.3+0.0}_{-0.0-2.9-0.0}$ \\
 $K_0^{*-}(1430)K^+$ & \qquad\quad $14.5^{+0.0+3.2+0.1}_{-0.0-2.9-0.1}$ ~~~ & $K_0^{*+}(1430)K^-$ & $1.0^{+0.0+1.0+0.0}_{-0.0-0.4-0.0}$ \\
 NR & \qquad\quad $23.8^{+0.2+9.9+0.0}_{-0.1-6.7
 -0.0}$ ~~~ & NR & $24.2^{+0.0+7.9+0.0}_{-0.0-5.1
 -0.0}$ \\ \hline
 Total  & \qquad\quad $35.3^{+0.3+15.7+0.0}_{-0.2-~9.8-0.0}$ ~~~ & Total  & $36.7^{+0.2+14.9+0.1}_{-0.2-~9.0-0.1}$  \\
\end{tabular}
\end{ruledtabular}
\end{table}

%%%%%%%%%%%%%%%%%%%%%%%%%%%%%%%%%%%%%%%%%%%%%%%%
\begin{figure}[t]
\centering
 \subfigure[]{
  \includegraphics[width=0.4\textwidth]{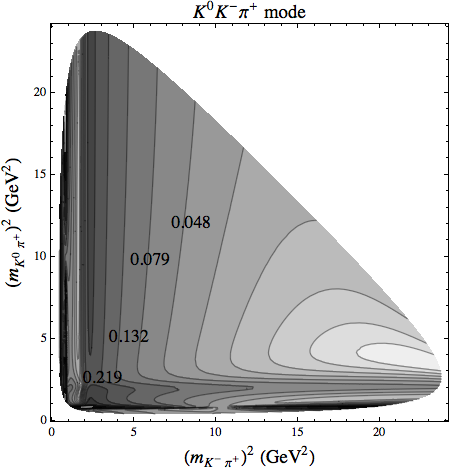}
} \hspace{0.2in} \subfigure[]{
  \includegraphics[width=0.4\textwidth]{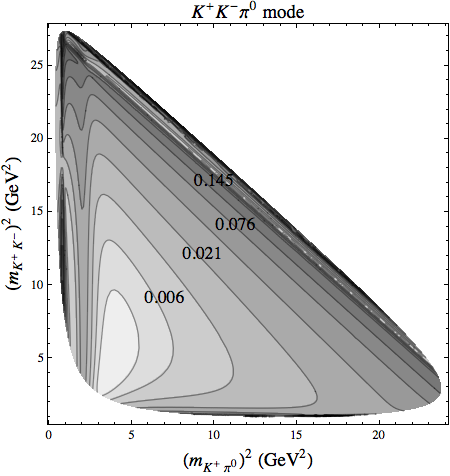}
}
\\ \subfigure[]{
  \includegraphics[width=0.4\textwidth]{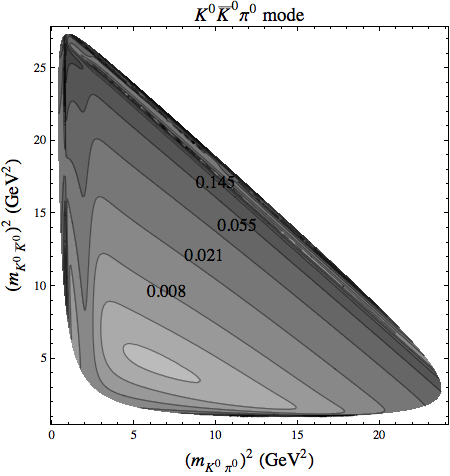}
}
\caption{Dalitz-plot distributions of penguin dominated $\overline B^0_s$ to (a) $K^0K^-\pi^+$,
(b) $K^+K^-\pi^0$ and (c)~$K^0\bar K^0\pi^0$ decays. The \CP averaged differential rates are in units of $10^{-6}$ GeV$^{-4}$.
Note that the contour lines are separated logarithmically.}
\label{fig:Dalitz1}
\end{figure}
%%%%%%%%%%%%%%%%%%%%%%%%%%%%%%%%%%%%%%%%%%%%%%%%

\subsection{Branching fractions}
We see from Table \ref{tab:BsBRCP} that the penguin-dominated modes $\Bs\to K^0 K^-\pi^+$ and $\ov K^0 K^+\pi^-$ have the largest rates among the three-body decays $\Bs\to PPP$ with $P=K,\pi$. Resonant and nonresonant contributions are shown in Table \ref{tab:BsKKpi}. \footnote{Although in $\Bs\to K_S K^\mp\pi^\pm$ decays, the rate of $\Bs\to K_0^{*+}(1430)K^-\to K^0K^-\pi^+$ is much smaller than that of $\Bs\to K_0^{*-}(1430)K^+\to \ov K^0K^+\pi^-$, it is the other way around in $\Bs\to \pi^0K^+K^-$ decays: $\B(\Bs\to K_0^{*+}(1430)K^-\to K^+K^-\pi^0)\gg
\B(\Bs\to K_0^{*-}(1430)K^+\to K^-K^+\pi^0)$.
Through the narrow width approximation,
 $\Gamma(B\to RP\to P_1P_2P)=\Gamma(B\to RP)\B(R\to P_1P_2)$
with $R$ being a resonance, it follows that the branching fraction of
$\Bs\to K_0^{*+}(1430)K^-$ is similar to that of $\Bs\to K_0^{*-}(1430)K^+$.
}
It is evident that they are dominated by the $\ov K_0^{*0}(1430)$, $K_0^{*-}(1430)$ scalar resonances and nonresonant contributions.
The Dalitz-plot distribution shown in Fig.~\ref{fig:Dalitz1}(a) also exhibits this feature.
For the nonresonant contributions, we find if the nonresonant matrix element of scalar density given by Eq. (\ref{eq:KpimeNew1})
is applied to the $B_s$ meson, the total branching fraction  of $\Bs\to \buildrel (-)\over {K^0} K^\mp\pi^\pm$ will be best accommodated with the parameter $\alpha\approx 0.10\,{\rm GeV}^{-2}$.

For the channel $\Bs\to K^0K^+K^-$, the current LHCb measurement of its branching fraction lies in the range $(0.2-3.4)\times 10^{-6}$, while our prediction $1.4\times 10^{-6}$ is in the middle. We see that both $\Bs\to K^+K^-\pi^0$ and $\Bs\to K^0\ov K^0\pi^0$ have very similar rates, as it should be. 

%The measured branching fraction of $\Bd\to \ov K^0\pi^+\pi^-$, $(50.2\pm1.5\pm1.8)\times 10^{-6}$ by BaBar \cite{BaBarK0pippim} and $(47.5\pm2.4\pm3.7)\times 10^{-6}$ by Belle \cite{BelleK0pipi}, is much larger than $\B(\Bs\to K^0K^+K^-)$. We see from the last  relation in Eq. (\ref{eq:Uspin}) that the branching fraction of $\Bs\to K^0K^+K^-$ is preferred to be close to the high end of the LHCb measurement; otherwise, the predicted \CP asymmetry of this mode will become too large.

%%%%%%%%%%%%%%%%%%%%%%%%%%%%%%%%%%%%%%%%%%%%%%%%
\begin{figure}[!]
\centering
\subfigure[]{
  \includegraphics[width=0.45\textwidth]{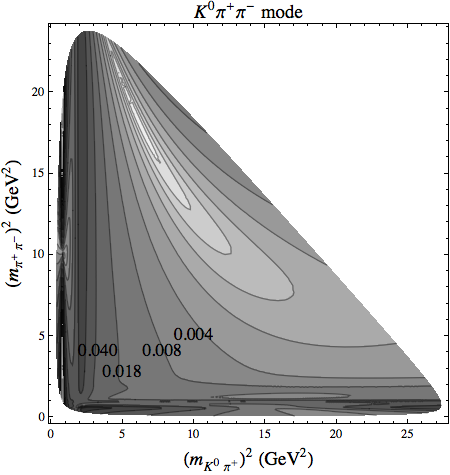}
} \hspace{0.2in} \subfigure[]{
  \includegraphics[width=0.45\textwidth]{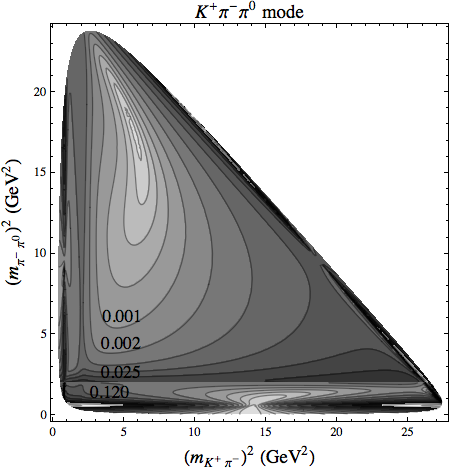}
}
\\\subfigure[]{
  \includegraphics[width=0.45\textwidth]{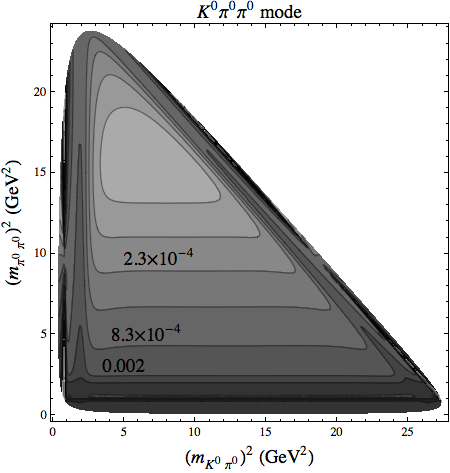}
} \hspace{0.2in} \subfigure[]{
  \includegraphics[width=0.45\textwidth]{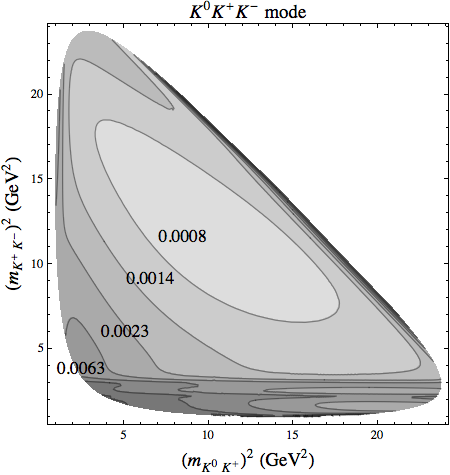}
}
\caption{Same as Fig.~\ref{fig:Dalitz1}, but for tree-dominated $\overline B^0_s$ to (a) $K^0\pi^+\pi^-$, (b) $K^+\pi^-\pi^0$,
(c) $K^0\pi^0\pi^0$ and (d) $K^0K^+K^-$ decays.
} \label{fig:Dalitz2}
\end{figure}
%%%%%%%%%%%%%%%%%%%%%%%%%%%%%%%%%%%%%%%%%%%%%%%%

As for the tree-dominated mode $\Bs\to K^0\pi^+\pi^-$, our prediction of its rate is consistent with experiment within errors.
See Fig.~\ref{fig:Dalitz2} for the Dalitz-plot distributions of tree-dominated modes.

\begin{table}[tb]
 \caption{Direct \CP asymmetries in $\Bs\to PPP$ decays via $U$-spin symmetry. Theoretical predictions of branching fractions of $\Bd\to \ov K^0 K^+ \pi^-$, $K^0 \pi^+K^-$  and \CP asymmetries for $\Bd\to PPP$ are taken from \cite{CC:3body}, while  branching fractions of $\Bd\to \ov K^0 K^+K^-$ and $\ov K^0 \pi^+\pi^-$ are quoted from \cite{BaBarK0pippim,BelleK0pipi} and \cite{BaBarKKK,BelleKKK,Aaij:Bs}, respectively.
 Only the central values are given for \CP asymmetries derived from $U$-spin symmetry.
 } \label{tab:BsUspin}
\begin{ruledtabular}
{\footnotesize
 \begin{tabular}{l c r |l c r c }
 {Modes} & $\B(10^{-6})$
   &  $\A_{CP}(\%)$ & Modes  & $\A_{CP}(\%)$($U$-spin) & $\A_{CP}(\%)$(theory) &  \\  \hline
   ${\overline{B}}^{0}_{d}\to \ov K^0 K^+ \pi^-$  \footnotemark[1]                    & $3.9^{+1.4}_{-0.9}$
        & $-9.4^{+1.7+1.5+0.0}_{-2.9-2.5-0.0}$
                                                                & ${\overline{B}}^{0}_{s}\to K^0K^-\pi^+$ &
                                                                $0.96$
                                                                & $4.6^{+0.3+1.1+0.1}_{-0.3-1.1-0.1}$
                                                                \\
  ${\overline{B}}^{0}_{d}\to K^0 K^- \pi^+$  \footnotemark[1]                    & $0.8^{+1.4}_{-0.9}$
        & $-12.7^{+1.7+1.5+0.0}_{-2.9-2.5-0.0}$
                                                                & ${\overline{B}}^{0}_{s}\to \ov K^0K^+\pi^-$ &
                                                                $0.28$
                                                                & $-1.9^{+0.1+0.1+0.0}_{-0.1-0.1-0.0}$
                                                                \\
    $\overline B^0_d\to \ov K^0 K^+ K^-$
       & $23.9\pm1.0$                                   & $-5.5^{+1.4+0.5+0.1}_{-1.0-0.5-0.1}$                                     & $\overline B^0_s\to K^0 \pi^+\pi^-$
                                                                 &
                                                                 $10.1$
                                                                 &
                                                                 $8.1^{+0.9+1.0+0.0}_{-0.0-0.9-0.0}$
                                                                 \\
   ${\overline{B}}^{0}_{d} \to \ov K^{0}\pi^+\pi^-$
     &$49.6\pm2.1$
     & $-0.8^{+0.0+0.1+0.0}_{-0.0-0.1-0.0}$
                                                                 & ${\overline{B}}^{0}_{s} \to K^{0}K^+K^-$
                                                                 &
                                                                 $15.9
                                                                 $
                                                                 & $-17.4^{+0.1+0.7+0.4}_{-0.2-0.5-0.4}$
                                                                 \\
   ${\overline{B}}^{0}_{d} \to K_SK_SK_S$
     & $6.19^{+1.62}_{-1.42}$
     & $0.7^{+0.0+0.0+0.0}_{-0.0-0.0-0.0}$
                                                                 & ${\overline{B}}^{0}_{s} \to K_SK_SK_S$
                                                                 &
                                                                 $-20.8$
                                                                 & $-13.4^{+0.1+0.4+0.2}_{-0.1-0.4-0.2}$
                                                                 \\
 \end{tabular}}
 \end{ruledtabular}
\footnotetext[1]{Branching fractions and \CP asymmetries of $\Bd\to \buildrel (-)\over {K^0} K^\pm\pi^\mp$ are calculated in \cite{CC:3body}, but not for the individual $\ov K^0 K^+\pi^-$ and $K^0 K^-\pi^+$ modes. Here we have corrected some typos in the computer code for $\Bd\to \buildrel (-)\over {K^0} K^\pm\pi^\mp$ and updated the results. }
 \end{table}

\subsection{Direct \CP asymmetries}
There are five $U$-spin relations exhibited in Eq. (\ref{eq:Uspin}). They cannot be tested at present due to the lack of experimental data.
Nevertheless, since the $U$-spin symmetry breaking is already included in the model calculations, we can check quantitatively how good the symmetry is. In Table \ref{tab:BsUspin} we show some of direct \CP asymmetries in $B_s$ decays evaluated using the $U$-spin relations Eq. (\ref{eq:Uspin}) and theoretical inputs for the branching fractions of $\ov B_{s}^0\to PPP$ decays and \CP asymmetries of $\Bd\to PPP$. We see that in general $\A_{CP}$ obtained by $U$-spin symmetry are not in good agreement with the direct calculation based on factorization. In particular, the signs of the calculated \CP rate asymmetries $\A_{CP}(\Bs\to K^0K^+K^-)$ and $\A_{CP}(\Bs\to \ov K^0K^+\pi^-)$ are opposite to what expected from the $U$-spin symmetry argument. This has to do with sizable $U$-spin symmetry breaking in some matrix elements. For example, the matrix element $\la \pi^-\ov K^0|\bar su|0\ra$ appearing in the decay amplitude of $\Bs\to\ov K^0K^+\pi^-$ should be the same as the one $\la K^-K^0|\bar du|0\ra$ in $\Bs\to K^0\pi^+K^-$ under $U$-spin symmetry. However, from Eq. (\ref{eq:KKssme}) we see that in reality $U$-spin symmetry which relates the matrix elements of $\la K^0K|\bar dq|0\ra$ to  $\la \ov K^0\pi|\bar sq|0\ra$ is badly broken. Therefore, it is not surprising that the $U$-spin relations are generally not well respected.

As mentioned in Introduction, LHCb has observed  interesting correlations of the \CP violation between the three-body decays of charged $B$ mesons given in Eq. (\ref{eq:CPcorrelation}).  It will be interesting to see experimentally if the analog relation $\A_{CP}(\Bs\to K^0\pi^+\pi^-)\approx -\A_{CP}(\Bs\to K^0K^+K^-)$ also holds in the $B_s$ sector. Our model calculation (see Table \ref{tab:BsBRCP}) indeed shows an opposite sign of \CP rate asymmetries between the above-mentioned two modes.

Besides inclusive \CP violation, LHCb has also observed large asymmetries in localized regions of phase space \cite{LHCb:Kppippim,LHCb:pippippim}. Again, it is important to check if significant signatures of \CP violation occur in three-body $B_s$ decays in the low mass regions devoid of most of the known resonances. If the phase space of the decay $\Bs\to K^+K^-\pi^0$ is restricted to the mass region $m^2_{K^+K^-}<1.5$ GeV$^2$ as it has been done for the LHCb measurement of the regional \CP violation in $B^-\to K^+K^-\pi^-$, we find $\A_{CP}^{\rm region}(\Bs\to K^+K^-\pi^0)=(29.6^{+0.5+1.4+0.1}_{-0.4-1.7-0.2})\%$ which is greater than the integrated inclusive \CP asymmetry of order $0.9\%$ (cf. Table \ref{tab:BsBRCP}).

\section{Conclusions}
We have presented in this work a study of charmless three-body $B_s$ decays within the framework of a simple model based on factorization. Our main results are:

\begin{itemize}

\item
The penguin-dominated decays $\Bs\to K^0 K^-\pi^+$ and $\ov K^0 K^+\pi^-$  have the largest rates among the three-body decays $\Bs\to PPP$ with $P$ being a kaon or pion, and they are dominated by the $K_0^*(1430)$ resonances and nonresonant contributions. The decays $\Bs\to K^+K^-\pi^0$ and $K^0\ov K^0\pi^0$ have similar rates and their branching fractions are of order $20\times 10^{-6}$.

\item
The branching fraction of $\Bs\to K^0K^+K^-$ is predicted to be $1.4\times 10^{-6}$, which is in the middle of the measured region  $(0.2-3.4)\times 10^{-6}$ obtained by LHCb. 

\item
Several model-independent relations based on $U$-spin symmetry are derived. Although they cannot be tested by the present available data, they can be checked by the dynamical calculations as shown in Table \ref{tab:BsUspin}. Because the $U$-spin symmetry which relates the matrix elements of scalar densities e.g. $\la K\ov K|\bar q_1q_2|0\ra$ and  $\la K\pi|\bar sq|0\ra$ is badly broken, $U$-spin symmetry relations for \CP violation in 3-body $B_s$ decays are generally not well respected.

\item
We found sizable \CP asymmetries in $K^0\pi^+\pi^-$, $K^0\pi^0\pi^0$, $K^0K^+K^-$ and $K_SK_SK_S$ channels. Just as the $B_u$ sector, the \CP asymmetries  of  $\Bs\to K^0\pi^+\pi^-$ and $\Bs\to K^0K^+K^-$ have similar magnitudes but are opposite in sign.

\end{itemize}

\vskip 2.0cm \acknowledgments
This research was supported in part by the National Center for Theoretical Sciences and the National Science
Council of R.O.C. under Grant Nos. NSC100-2112-M-001-009-MY3 and NSC100-2112-M-033-001-MY3.

%%%%%%%%%%%%%%%%%%%%
\appendix
%%%%%%%%%%%%%%%%%%%

\section{Factorizable amplitudes of $B_s\to PPP$ decays}

In the following we list the factorizable amplitudes of various $\Bs\to PPP$ decays:

\be \label{eq:AK0Kppim}
 \la {K^0} K^-\pi^+  |T_p|\ov B^0_s\ra &=&
 \la \pi^+ K^0|(\bar u b)_{V-A}|\ov B^0_s\ra \la K^-|(\bar s u)_{V-A}|0\ra
 \left[a_1 \delta_{pu}+a^p_4+a_{10}^p-(a^p_6+a^p_8) r_\chi^K \right]
 \non\\
 &&+\la K^0|(\bar d b)_{V-A}|\ov B_s^0\ra
       \la \pi^+K^-|(\bar s d)_{V-A}|0\ra
       \left[a^p_4-{1\over 2}a^p_{10}\right]
       \non\\
 &&+\la K^0|\bar d b|\ov B_s^0\ra
       \la \pi^+K^-|\bar s d|0\ra
       (-2 a^p_6+a^p_8)
       \non\\
  &&  +\la {K^0} K^-\pi^+|(\bar uu)_{V-A}|0\ra
     \la 0|(\bar s b)_{V-A}|\ov B^0_s\ra
       \bigg(a_2\delta_{pu}+a_3+a_9\bigg)
       \non\\
   &&  +\la {K^0} K^-\pi^+|(\bar uu)_{V+A}|0\ra
     \la 0|(\bar s b)_{V-A}|\ov B^0_s\ra
       (a_5+a_7)
       \non\\
 &&  + \la {K^0} K^-\pi^+|\bar s(1+\gamma_5) d|0\ra
       \la 0|\bar s\gamma_5 b|\ov B_s^0\ra
       (2a^p_6-a^p_8),
\en

\be \label{eq:AK0Kmpip}
 \la \ov K^0 K^+\pi^-  |T_p|\ov B^0_s\ra &=&
 \la \pi^-K^+|(\bar d b)_{V-A}|\ov B^0_s\ra
       \la \ov K^0|(\bar sd)_{V-A}|0\ra  \left[a^p_4-{1\over 2}a^p_{10}-(a^p_6-{1\over 2}a^p_8) r_\chi^K\right]
       \non\\
 &&+\la K^+|(\bar u b)_{V-A}|\ov B_s ^0\ra
                   \la \pi^- \ov K^0|(\bar s u)_{V-A}|0\ra
    \left[a_1\delta_{pu}+a_4^p+a_{10}^p \right]
                   \non\\
 &&+\la K^+|\bar u b|\ov B_s ^0\ra
   \la \pi^- \ov K^0|\bar s u|0\ra (-2a^p_6-2a^p_8)
                   \non\\
  &&  +\la \ov K^0 K^+\pi^-|(\bar uu)_{V-A}|0\ra
     \la 0|(\bar s b)_{V-A}|\ov B^0_s\ra
       \bigg(a_2\delta_{pu}+a_3+a_9\bigg)
       \non\\
   &&  +\la \ov K^0 K^+\pi^-|(\bar uu)_{V+A}|0\ra
     \la 0|(\bar s b)_{V-A}|\ov B^0_s\ra
       (a_5+a_7)
       \non\\
 &&  + \la \ov K^0 K^+\pi^-|\bar s(1+\gamma_5) d|0\ra
       \la 0|\bar s\gamma_5 b|\ov B_s^0\ra
       (2a^p_6-a^p_8),
\en

\be \label{eq:pi0KpKm}
 \la \pi^0 K^+ K^-|T_p|\ov B^0_s\ra &=&
 \la K^+ \pi^0|(\bar u b)_{V-A}|\ov B^0_s\ra \la K^-|(\bar s u)_{V-A}|0\ra
 \left[a_1 \delta_{pu}+a^p_4+a_{10}^p-(a^p_6+a^p_8) r_\chi^K\right]
 \non\\
  &&+  \la K^+ K^-|(\bar s b)_{V-A}|\ov B_s^0\ra \la \pi^0|(\bar u u)_{V-A}|0\ra
 \left[a_2\delta_{pu}+a_3-a_5-a_7+a_9\right] \non \\
  &&+  \la K^+ K^-|(\bar s b)_{V-A}|\ov B_s^0\ra \la \pi^0|(\bar d d)_{V-A}|0\ra
 \left[ a_3-a_5+{1\over 2}(a_7-a_9)\right] \non \\
 &&+
 \la K^+ |(\bar u b)_{V-A}|\ov B_s^0\ra \la K^-\pi^0|(\bar s u)_{V-A}|0\ra
 \left[a_1 \delta_{pu}+a^p_4+a_{10}^p\right] \non \\
 &&+
 \la K^+ |\bar u b|\ov B_s^0\ra \la K^-\pi^0|\bar s u|0\ra
 (-2a^p_6-2a_{8}^p)  \\
 &&+ \la \pi^0 K^+K^-|(\bar ss)_{V-A}|0\ra\la 0|(\bar sb)_{V-A}|\Bs\ra \left[a_3+a_4^p-{1\over 2}(a_9+a_{10}^p)\right] \non \\
  &&+ \la \pi^0 K^+K^-|(\bar ss)_{V+A}|0\ra\la 0|(\bar sb)_{V-A}|\Bs\ra \left[a_5-{1\over 2}a_7\right], \non
\en

 \be  \label{eq:AK0pippim}
 \la K {}^0 \pi^+ \pi^-|T_p|\ov B^0_s\ra&=&
 \la \pi^+ K {}^0|(\bar u b)_{V-A}|\ov B {}^0_s\ra \la \pi^-|(\bar d u)_{V-A}|0\ra
 \left[a_1 \delta_{pu}+a^p_4+a_{10}^p-(a^p_6+a^p_8) r_\chi^\pi \right]
 \non\\
 &&+ \la \pi^+ \pi^-|(\bar s b)_{V-A}|\ov B {}^0_s\ra \la K^0|(\bar d s)_{V-A}|0\ra
 \left[a^p_4-{1\over 2}a_{10}^p- (a_6^p-{1\over 2}a_8^p) r_\chi^K \right]
 \non\\
 &&+\la K {}^0|(\bar d b)_{V-A}|\ov B {}^0_s\ra
                   \la \pi^+ \pi^-|(\bar u u)_{V-A}|0\ra
    (a_2\delta_{pu}+a_3+a_5+a_7+a_9)
                   \non\\
  &&+\la K {}^0|(\bar d b)_{V-A}|\ov B {}_s^0\ra
                   \la \pi^+ \pi^-|(\bar d d)_{V-A}|0\ra
    \bigg[a_3+a^p_4+a_5-\frac{1}{2}(a_7+a_9+a^p_{10})\bigg]
    \non\\
 &&+\la K {}^0|(\bar d b)_{V-A}|\ov B {}_s^0\ra
                   \la \pi^+ \pi^-|(\bar s s)_{V-A}|0\ra
    \bigg[a_3+a_5-\frac{1}{2}(a_7+a_9)\bigg]
    \non\\
 &&+\la K {}^0|\bar d b|\ov B {}^0_s\ra
       \la \pi^+ \pi^-|\bar dd|0\ra
       (-2 a^p_6+a^p_8)
       \non\\
  &&  +\la \pi^+ \pi^-K {}^0|(\bar d s)_{V-A}|0\ra
     \la 0|(\bar s b)_{V-A}|\ov B {}_s^0\ra
       \left[a^p_4-\frac{1}{2} a^p_{10}\right]
       \non\\
 &&  + \la \pi^+ \pi^- K {}^0|\bar d(1+\gamma_5) s|0\ra
       \la 0|\bar s\gamma_5 b|\ov B {}_s^0\ra
       (2a^p_6-a^p_8),
 \en

 \be \label{eq:AK0KpKm}
 \la K^0 K^+ K^-|T_p|\ov B^0_s\ra &=&
 \la K^+ K^-|(\bar s b)_{V-A}|\ov B_s^0\ra \la K^0|(\bar d s)_{V-A}|0\ra
 \left[a^p_4-{1\over 2}a_{10}^p-(a^p_6-{1\over 2}a^p_8) r_\chi^K\right]
 \non\\
&& + \la  K^0|(\bar db)_{V-A}|\ov B_s^0\ra \la K^+K^-|(\bar
uu)_{V-A}|0\ra\left[a_2\delta_{pu}+a_3+a_5+a_7+a_9)\right] \non \\
&& + \la  K^0|(\bar db)_{V-A}|\ov B_s^0\ra \la K^+K^-|(\bar
dd)_{V-A}|0\ra\left[a_3+a_4^p+a_5-{1\over 2}(a_7+a_9+a_{10}^p)\right] \non \\
&& + \la  K^0|(\bar db)_{V-A}|\ov B_s^0\ra \la K^+K^-|(\bar
ss)_{V-A}|0\ra\left[a_3+a_5-{1\over 2}(a_7+a_9)\right] \non \\
&& +\la K^+|(\bar ub)_{V-A}|\ov B_s^0\ra \la K^0K^-|(\bar
du)_{V-A}|0\ra \left[a_1+a_4^p+a_{10}^p \right]\non \\
&& +\la K^0|\bar db|\ov B_s^0\ra \la K^+K^-|\bar dd|0\ra
(-2a_6^p+a_8^p)  \\
&& +\la K^+|\bar ub|\ov B_s^0\ra \la K^0K^-|\bar
du|0\ra (-2a_6^p-2a_8^p)\non \\
&& +\la K^0 K^+ K^-|(\bar ds)_{V-A}|0\ra \la0|(\bar
sb)_{V-A}|\ov B_s^0\ra \left[a_4^p-{1\over 2}a_{10}^p\right] \non \\
&& +\la K^0 K^+K^-|\bar d(1+\gamma_5)s|0\ra \la 0|\bar
s\gamma_5b|\ov B_s^0\ra(2a_6^p-a_8^p), \non
 \en

 \be \label{eq:AK0K0K0}
 \la \ov K^0  K
 {}^0 K {}^0|T_p|\ov B^0_s\ra &=&
 \la \ov K^0 K {}^0|(\bar s b)_{V-A}|\ov B {}_s^0\ra \la K {}^0|(\bar d s)_{V-A}|0\ra
 \Big(a^p_4-\frac{1}{2}a^p_{10}-(a^p_6-\frac{1}{2}a^p_8)
 r_\chi^K\Big)
 \non\\
         &&+\la K {}^0 |(\bar d b)_{V-A}|\ov B {}_s^0\ra
       \la K^0 \ov K {}^0|(\bar u u)_{V-A}|0\ra
 \left[a_2\delta_{pu}+a_3+a_5+a_7+a_9)\right] \non\\
       &&+\la K {}^0 |(\bar d b)_{V-A}|\ov B {}_s^0\ra
       \la K^0 \ov K {}^0|(\bar d d)_{V-A}|0\ra
 \left[a_3+a^p_4+a_5-\frac{1}{2}(a_7+a_9+a_{10}^p)\right] \non\\
        &&+\la K {}^0 |(\bar d b)_{V-A}|\ov B {}_s^0\ra
       \la K^0 \ov K {}^0|(\bar s s)_{V-A}|0\ra
 \left[a_3+a_5-\frac{1}{2}(a_7+a_9)\right] \non\\
  &&+\la K {}^0 |\bar d b|\ov B {}_s^0\ra
       \la K^0 \ov K {}^0|\bar d d|0\ra
       (-2 a^p_6+a^p_8)
       \non\\
   &&  +   \la \ov K^0  K {}^0  K {}^0|(\bar d s)_{V-A}|0\ra
      \la 0|(\bar sb)_{V-A}|\ov B {}_s^0\ra
       \left[a_4^p-{1\over 2}a_{10}^p\right] \non \\
  &&  +   \la \ov K^0 K {}^0  K {}^0|\bar d(1+\gamma_5) s|0\ra
      \la 0|\bar s\gamma_5 b|\ov B {}_s^0\ra
       (2a^p_6-a^p_8),
 \en

 \be \label{eq:AKppimpi0}
 \la K^+\pi^- \pi^0|T_p|\Bs\ra &=&
 \la K^+ \pi^0|(\bar u b)_{V-A}|\ov B_s^0\ra \la \pi^-|(\bar d u)_{V-A}|0\ra
 \left[a_1\delta_{pu}+a^p_4+a_{10}^p-(a^p_6+a^p_8) r_\chi^\pi\right]
 \non \\
 && + \la \pi^-K^+|(\bar
db)_{V-A}|\ov B_s^0\ra\la \pi^0|(\bar uu)_{V-A}|0\ra
\left[a_2\delta_{pu}+a_3-a_5-a_7+a_9 \right] \non \\
 && + \la \pi^-K^+|(\bar
db)_{V-A}|\ov B_s^0\ra\la \pi^0|(\bar dd)_{V-A}|0\ra
\left[a_3-a_5+{1\over 2}(a_7-a_9) \right] \non \\
 &&  +\la K^+ |(\bar u b)_{V-A}|\ov B_s^0\ra \la \pi^-\pi^0|(\bar d
u)_{V-A}|0\ra \left[a_1\delta_{pu}+a^p_4+a_{10}^p\right]
 \non \\
 &&  +\la K^+ |\bar u b|\ov B_s^0\ra \la \pi^-\pi^0|\bar d
u|0\ra (-2a^p_6-2a_{8}^p)
 \non \\
&& +\la K^+\pi^-\pi^0|(\bar ds)_{V-A}|0\ra \la 0|(\bar
sb)_{V-A}|\ov B_s^0\ra \left[a_4^p-{1\over 2}a_{10}^p\right] \non \\
&& +\la K^+\pi^-\pi^0|\bar d(1+\gamma_5)s|0\ra \la 0|\bar
s\gamma_5b|\ov B_s^0\ra(2a_6^p-a_8^p),
 \en

\be \label{eq:AK0pi0pi0}
 \la K^0 \pi^0 \pi^0|T_p|\Bs \ra &=&
 \la K^0\pi^0|(\bar db)_{V-A}|\Bs \ra\la \pi^0|(\bar uu)_{V-A}|0\ra
\left[a_2\delta_{pu} +a_3-a_5-a_7+a_9\right] \non \\
 && \la K^0\pi^0|(\bar db)_{V-A}|\Bs \ra\la \pi^0|(\bar dd)_{V-A}|0\ra
\left[a_3-a_5+{1\over 2}(a_7-a_9)\right] \non \\
 && +\la\pi^0\pi^0|(\bar uu)_\vma|0\ra \la K^0|(\bar
 db)_\vma|\Bs \ra \left[a_2\delta_{pu}+a_3+a_5+a_7+a_9\right] \non \\
 && +\la\pi^0\pi^0|(\bar dd)_\vma|0\ra \la K^0|(\bar
 db)_\vma|\Bs \ra \left[a_3+a_4^p+a_5-{1\over
 2}(a_7+a_9+a_{10}^p)\right] \non \\
 && + \la\pi^0\pi^0|\bar dd|0\ra\la K^0|\bar db|\Bs \ra(-2a_6^p+a_8^p) \non \\
&& +\la K^0\pi^0\pi^0|(\bar ds)_{V-A}|0\ra \la 0|(\bar
sb)_{V-A}|\Bs \ra \left[a_4^p-{1\over 2}a_{10}^p\right] \non \\
&& +\la K^0\pi^0\pi^0|\bar d(1+\gamma_5)s|0\ra \la0|\bar
s\gamma_5b|\Bs \ra(2a_6^p-a_8^p),
 \en

\be \label{eq:AK0K0pi0}
 \la  \pi^0 K^0 \ov K^0|T_p|\Bs \ra &=&
 \la K^0\ov K^0|(\bar sb)_{V-A}|\Bs \ra\la \pi^0|(\bar uu)_{V-A}|0\ra
\left[a_2\delta_{pu} +a_3-a_5-a_7+a_9\right] \non \\
&&  +\la K^0\ov K^0|(\bar sb)_{V-A}|\Bs \ra\la \pi^0|(\bar dd)_{V-A}|0\ra
\left[a_3-a_5+{1\over 2}(a_7-a_9)\right] \non \\
 && + \la \ov K^0|(\bar sd)_{V-A}|0\ra\la K^0\pi^0|(\bar db)_{V-A}|\Bs \ra\left[a_4^p-{1\over 2}a^p_{10}-(a_6^p-{1\over 2}a_8^p)r_\chi^K\right] \non \\
  && + \la \ov K^0\pi^0|(\bar sd)_{V-A}|0\ra\la K^0|(\bar db)_{V-A}|\Bs \ra\left[a_4^p-{1\over 2}a^p_{10}\right] \non \\
  && + \la \ov K^0\pi^0|\bar sd|0\ra\la K^0|\bar db|\Bs \ra(-2a_6^p+a_8^p) \non \\
 &&+ \la \pi^0 K^0\ov K^0|(\bar ss)_{V-A}|0\ra\la 0|(\bar sb)_{V-A}|\Bs\ra \left[a_3+a_4^p-{1\over 2}(a_9+a_{10}^p)\right] \non \\
  &&+ \la \pi^0 K^0\ov K^0|(\bar ss)_{V+A}|0\ra\la 0|(\bar sb)_{V-A}|\Bs\ra \left[a_5-{1\over 2}a_7\right]. \non
 \en

%%%%%%%%%%%%%%%%%%%%%%%%%%%%%%%%%%%%%%%%%%%%%%%%%%%%%%%%

\end{document}